\documentclass[showpacs,floatfix,nofootinbib]{revtex4-1}
\usepackage{graphicx,psfrag,amsmath,amssymb,amsfonts,latexsym,color,dcolumn}
\begin{document}
\title{The derivative expansion approach to the interaction between
close surfaces}
\author{C. D. Fosco$^a$}
\author{F. C. Lombardo$^b$}
\author{F. D. Mazzitelli$^{a}$}
\affiliation{$^a$Centro At\'omico Bariloche and Instituto Balseiro,
 Comisi\'on Nacional de Energ\'\i a At\'omica, \\
R8402AGP Bariloche, Argentina}
\affiliation{$^b$Departamento de F\'\i sica {\it Juan Jos\'e
Giambiagi}, FCEyN UBA and IFIBA CONICET-UBA, Facultad de Ciencias Exactas y Naturales,
Ciudad Universitaria, Pabell\' on I, 1428 Buenos Aires, Argentina}
\date{today}
\begin{abstract} 
The derivative expansion approach to the calculation of the
interaction between two surfaces, is a generalization of the proximity
force approximation, a technique of widespread use in different areas
of physics.  The derivative expansion has so far been applied to seemingly unrelated problems
in different areas; it is our principal aim here to present the approach in
its full generality. To that end, we introduce an unified setting, which is
independent of any particular application, provide a formal derivation of
the derivative expansion in that general setting, and study some its properties.  With a view
on the possible application of the derivative expansion to other areas, like nuclear and
colloidal physics, we  also discuss the relation between the derivative expansion and some
time-honoured uncontrolled approximations used in those contexts.  By
putting them under similar terms as the derivative expansion, we believe that the path is
open to the calculation of next to leading order corrections also
for those contexts.  We also review some results obtained within the derivative expansion, by
applying it to different concrete examples and highlighting some important
points. 
\end{abstract}
\maketitle
\section{Introduction}\label{sec:intro}
The Proximity Force Approximation  (PFA) has been widely used in many areas of physics, as a tool to compute
the total force between smooth surfaces at short distances, the
scale of that smoothness being set by the distance between them. Although that 
total force may result from very different underlying
microscopic mechanisms, the PFA is essentially the same in all of them,
since its basis is geometrical.

The main idea behind this approximation was introduced by Derjaguin in
1934~\cite{Derjaguin34}, when studying the effect of contact deformations 
on the adhesion of particles.  In that context, the so called Derjaguin
Approximation (DA) has, as its main ingredient, the  (assumed) knowledge of 
$E_{\parallel}(h)$, the interaction energy per unit area for 
two (infinite) plane parallel surfaces separated by a distance $h$.
The DA tells us that then the interaction energy $U_{\rm DA}(d)$ between two {\em
curved\/} surfaces is~\cite{Derjaguin34,books}
\begin{equation}
U_{\rm DA}(d)=2\pi R_{\rm eff}\int_d^\infty E_{\parallel}(h)dh\, ,
\label{DA}
\end{equation}  
where  $d$ is the distance between the surfaces, $R_1$ and $R_2$ their
respective curvature radii evaluated at the point of closest approach, and
$R_{\rm eff}=R_1R_2/(R_1+R_2)$.  The same approximation can be alternatively
written in terms of the force, $f_{\rm DA}$, as follows:
\begin{equation}
f_{\rm DA}(d)=2\pi R_{\rm eff} E_{\parallel}(d)\, .
\label{DAforce}
\end{equation}  
The usual derivation of this approximation relies upon the rather reasonable 
assumption that the interaction energy between the surfaces can be approximated 
by the PFA expression:
\begin{equation}
U_{\rm PFA}=\int \, dS\, E_{\parallel}\; ,
\label{PFA}
\end{equation}
where the integration is performed over just one of the surfaces, or even
over an intermediate mathematical surface lying  between the two physical
surfaces~\footnote{When applied to a compact object in front of a plane,
the integration is usually restricted to the portion of the compact
object's surface that faces the plane.}. Assuming that the surfaces are
gently curved, and approximating them by portions of the osculating spheres of radii
$R_1$ and $R_2$ at the point of closest approach, one arrives at the
DA.

The DA has been used to compute forces in many different physical
situations: colloidal and macromolecular phenomena, nuclear physics,
electrostatic forces, van der Waals interactions, Casimir forces, etc.  The
DA has been generally assumed to be an uncontrolled approximation,
presumably working well for close and gently curved surfaces. Its major
drawback was, perhaps, the absence of a procedure to asses the importance of the
next to leading order (NTLO)  corrections since, by construction, the DA is
not obtained as the leading term of any expansion.  In spite of that drawback,
surprisingly few works have been devoted to implement a systematic
improvement of the DA, which could give the PFA a more solid foundation,
and a way to improve it. 

Recently, we presented a new approximation scheme, the so-called Derivative
Expansion (DE), originally introduced within the context of the Casimir
effect, for the calculation of the interaction energy between two surfaces,
one of them flat and the other slightly curved~\cite{Foscoetal2011,Fosco:2012gp}.  This
approximation has been shown to be a natural extension of the PFA, and it has proven to be useful in rather
different situations, not just for Casimir effect calculations. The DE
approach provides a systematic way of introducing the DA and the PFA, and
under some circumstances, also of evaluating the NTLO corrections. 	

In this paper, our principal aim is to formulate and derive the DE in a quite general
form, so that previous (and hopefully new) applications of it may be
regarded as particular cases.  
To that end, we shall present a derivation of the PFA and its NTLO correction from first principles,
for the particular case of a curved surface described by a function
$x_3=\psi(x_1,x_2)$ in front of a plane at $x_3=0$. We shall also interpret the result
of this derivation in physical/geometrical terms.

The surfaces involved shall have quite different
physical realizations, depending on the system considered. Indeed, in
some examples they may correspond to two physical objects with very
small widths, interacting as a result of the presence of electric charge or
dipole layers on them.  In other cases, they may instead correspond
to {\em interfaces\/} between different material media.  Besides, the
nature of the `microscopic' interaction that produces the interaction may
also have quite different origins.  For example, it may be electrostatic,
mediated by a short-range force, or even be the result of a more involved
phenomenon, like the Casimir effect.  As it stems from the different
nature of the examples mentioned above, no assumption will be made about
whether the interaction between the surfaces proceeds from a microscopic
interaction that satisfies a superposition principle or not.

After presenting the general derivation, we shall make contact with other
efforts made in the literature towards generalizing and improving the DA,
like the Surface Element Integration (SEI)~\cite{SEI}, or the
Surface Integration Approach (SIA)~\cite{SIA}
introduced in the context of colloidal physics, as well as the different
PFA�s used in nuclear physics~\cite{Blocki1,Blocki2}.

This paper is organized as follows: in Section~\ref{sec:derivation} we
provide a first principles derivation of the DE, and then a construction of 
it using physical arguments. We also consider some properties of
the general formulae so obtained, and comment on the form of the higher
order terms.
Then in Section~\ref{sec:other} we discuss, from the point of view of the DE, some
generalizations of the DA that have been proposed in the literature, mostly
in the context of colloidal and nuclear physics. Based on a formal analogy
with the Casimir interaction  between almost transparent media,  we  make
contact with the SEI approach used in colloidal physics.  In
Section~\ref{sec:appl} we briefly review results obtained during the last
years using the DE, along with discussions of particular examples.
Finally, Section~\ref{sec:concl} contains our conclusions.

\section{Derivation and properties of the Derivative Expansion}\label{sec:derivation}

\subsection{Derivation by resummation of a perturbative expansion}\label{ssec:demo}

Let us first set up the problem: regardless of the interaction considered,
the geometry of the system shall be assumed to be as follows: one of the
surfaces, $L$, will be a plane, which (by a proper choice of coordinates)
shall be described by the equation $x_3 = 0$. Denoting by $x_1$, $x_2$ and
$x_3$ the set of three orthogonal Cartesian coordinates, the other surface,
$R$, is assumed to be describable by using a single Monge patch. Namely,
that it can be parametrized in terms of just a single function
$\psi({\mathbf x_\parallel})$, with \mbox{${\mathbf x_\parallel} =
(x_1,x_2)$}, such that $x_3 = \psi({\mathbf x_\parallel})$.  

To begin with, we note that the
interaction between the two surfaces shall be a functional $F[\psi]$, of a
single function $\psi({\mathbf x_\parallel})$. That functional may be an
energy, free energy, force, etc., depending on the context and the kind of system
being considered. 

The PFA approximation to $F$, which will be denoted here by $F_0$, 
is obtained by adding, for each point ${\mathbf x_\parallel}$,
the product of a local surface density ${\mathcal F}_0(\psi({\mathbf
x_\parallel}))$ depending only on the value of $\psi$ at the point
${\mathbf x_\parallel}$, times the surface element area; namely,
\begin{equation}
F_0 [\psi]\;=\; \int d^2{\mathbf x_\parallel}
\, {\mathcal F}_0(\psi({\mathbf x_\parallel})) \;. 
\end{equation} 
The local surface density is in turn determined by the knowledge of the
exact form of $F$ for the case of two parallel surfaces, in the following
way:
\begin{equation}
{\mathcal F}_0(a) \;=\; \lim_{{\mathcal S} \to \infty} 
\big[\frac{F[a]}{\mathcal S} \big]\;, 
\end{equation} 
where ${\mathcal S}$ denotes the area of the $L$ plate.

Namely, to determine the density one only needs to know the functional $F$
for  constant functions $\psi \equiv a$, and then extract a
surface factor due to translation invariance. That density is then
evaluated at the local distance between plates, $\psi({\mathbf
x_\parallel})$, multiplied by $d^2{\mathbf x_\parallel}$, and integrated.
When the functional $F$ describes the interaction energy between 
the surfaces, the local density  ${\mathcal F}_0$ is just the interaction 
energy per unit area $E_{\parallel}$, and $F_0$ becomes  $U_{PFA}$
(see Eq.(\ref{PFA})).

We will now show that one can derive the PFA and its corrections
by considering 
\begin{equation}
\psi({\mathbf x_\parallel}) = a + \eta( {\mathbf
x_\parallel})\, ,
\end{equation} 
and performing a resummation of the perturbative
expansion in powers of $\eta$. We start by expanding the functional $F[\psi]$ 
in powers of $\eta$: 
\begin{equation}
F[\psi]={\mathcal S\mathcal F_0}(a)+\sum_{n\geq 1}\int \frac{d^2k_\parallel^{(1)}}{(2\pi)^2}
...\frac{d^2k_\parallel^{(n)}}{(2\pi)^2} \,
\delta(k_\parallel^{(1)}+...+k_\parallel^{(n)}) \, h^{(n)}(k_\parallel^{(1)},...,k_\parallel^{(n)})
\, {\widetilde{\eta}}(k^{(1)}_\parallel) ...{\widetilde{\eta}}(k^{(n)}_\parallel)\; ,
\label{genexp}
\end{equation}
where the form factors $h^{(n)}$ could be computed by using standard perturbative
techniques \cite{4thorder}. They may depend on $a$, although we do not write explicitly
this dependence in order to simplify the notation. 

Now we see that, for a smooth function $\eta$, the Fourier transform $\widetilde\eta$
will be peaked at zero momentum; therefore, inside Eq.(\ref{genexp}) we can
approximate:
$h^{(n)}(k_\parallel^{(1)},...,k_\parallel^{(n)})\simeq h^{(n)}(0,...,0)$.  
As a consequence:
\begin{equation}
F(\psi)\simeq {\mathcal S \mathcal F_0}(a)+\sum_{n\geq
1}h^{(n)}(0,..,0)\int d^2{\mathbf x_\parallel} \eta({\mathbf
x_\parallel})^n \;.
\label{low}
\end{equation}
In principle, one could evaluate the form factors at zero momentum explicitly.
However, there is a shortcut that allows one to obtain all of them at once: for a
constant $\eta({\mathbf x_\parallel})=\eta_0$, the interaction energy is simply
given by Eq.(\ref{low}) with the replacement $\int
d^2{\mathbf x_\parallel} \,\eta({\mathbf x_\parallel})^n\rightarrow
{\mathcal S}\eta_0^n$. 
But for this particular case, $F$ is just the functional for parallel
plates separated by a distance $a+\eta_0$, namely,
\begin{equation}
{\mathcal F_0}(a+\eta_0)= {\mathcal F_0}(a) + \sum_{n\geq 1}h^{(n)}(0,..,0) \eta_0^n\, .
\label{F_0fin}
\end{equation}
Therefore, in this low-momentum approximation, the perturbative series can be
summed up, the result being: 
\begin{equation}
F_0[\psi]\simeq  
\int d^2{\mathbf x_\parallel} {\mathcal F_0}\left (a+\eta({\mathbf x_\parallel})\right)= 
\int d^2{\mathbf x_\parallel} {\mathcal F_0}(\psi)\, ,
\end{equation} 
which is just the PFA.

The straightforward calculation above has shown that, for the class of geometries
considered in this paper, the PFA can be derived from first principles
by performing a resummation of the perturbative calculation for the case of almost flat
surfaces. The PFA will be well defined as long as the form factors
$h^{(n)}(k_\parallel^{(1)},...,k_\parallel^{(n)})$
have a finite limit as $k_\parallel^{(i)}\to 0$.

This procedure also suggests that the PFA can be improved by considering
the NTLO in the expansion of the form factors.  We will assume that the
form factors can be expanded in powers of the momenta up to order two. This
is a nontrivial assumption: depending of the details of the interaction
considered, the low momentum behavior of the form factors could include non-analyticities. 
When that is not the case, we can perform the following expansion:
\begin{equation}
 h^{(n)}(k_\parallel^{(1)},...,k_\parallel^{(n)})= h^{(n)}(0,...,0) +
\sum_{i,\alpha}A^{(n)}_{i\alpha}k_{\parallel\, \alpha}^{(i)}+
\sum_{i,j,\alpha,\beta}B^{(n)}_{ij\alpha\beta}
 k_{\parallel\, \alpha}^{(i)}  k_{\parallel\, \beta}^{(j)}\,\ldots , 
\end{equation}
 for some $a-$dependent coefficients $A^{(n)}_{i\alpha}$ and
$B^{(n)}_{ij\alpha\beta}$. Here $i,j=1,...,n$ denote the different
arguments of the form factor and $\alpha,\beta = 1,2 $ their components.
Symmetry considerations allow us to simplify the above expression. Indeed,
rotational invariance implies that the form factors depend only
 on the scalar products $k_\parallel^{(i)}\cdot k_\parallel^{(j)}$.
Moreover, they must be symmetric under the interchange of any two momenta. As a
consequence 
\begin{equation}
 h^{(n)}(k_\parallel^{(1)},...,k_\parallel^{(n)})= h^{(n)}(0,...,0) + B^{(n)}\sum_i k_\parallel^{(i)\, 2}+ C^{(n)}\sum_{i\neq j}k_\parallel^{(i)}\cdot k_\parallel^{(j)} \, ,
 \label{cuad}
  \end{equation}
for some coefficients $B^{(n)}$ and $C^{(n)}$. Inserting Eq.\eqref{cuad}
into Eq.\eqref{genexp}, and performing integrations by parts, 
we find the following correction to the PFA
\begin{equation}
F_2[\psi]=\int d^2{\mathbf x_\parallel}\, \left[\sum_{n\geq 2}D^{(n)}\,
\eta^{n-2}\right] |\nabla \eta |^2\;,
\label{orden2}
\end{equation}
where the coefficients $D^{(n)}$ are linear combinations of $B^{(n)}$ and $C^{(n)}$. The subindex  $2$ indicates the number of derivatives.

To complete the calculation, the next step is to calculate the sum in
Eq.~\eqref{orden2}. This can be done by evaluating the correction $F_2$ for 
the particular case $\eta({\mathbf x_\parallel})= \eta_0 + \epsilon({\mathbf x_\parallel})$, with $\epsilon\ll\eta_0$, and
expanding up to second order in $\epsilon$. For this particular case
\begin{equation}
F_2[a+\eta_0+\epsilon]=\int d^2{\mathbf x_\parallel}\, \left[\sum_{n\geq
2}D^{(n)}\, \eta_0^{n-2}\right] |\nabla \epsilon |^2\, .
\label{orden2eps}
\end{equation}
Once more, the resummation can be performed, in this case by considering
the usual perturbative evaluation of the interaction energy up to second
order in $\epsilon$. This calculation will depend on the particular interaction considered, and from the result 
one can obtain the series above, that we will denote by $Z$. More explicitly
\begin{equation}
Z(a+\eta_0)\equiv \sum_{n\geq 2}D^{(n)}\, \eta_0^{n-2}\, .
\label{zeta2}
\end{equation}
Making the replacement $\eta_0\to\eta$ in the above equation, we arrive at
\begin{equation}
F_2[\psi]=\int d^2{\mathbf x_\parallel}\, Z(\psi) |\nabla \psi |^2\, ,
\label{fin}
\end{equation}
which is the NTLO correction to the PFA. 
This concludes the derivation of the PFA and its first correction, that reads
\begin{equation}\label{eq:de}
F_{\rm DE}[\psi] \;=\; \int d^2{\mathbf x_\parallel} \, \Big[ V(\phi) \, + \,
 Z(\psi) |\nabla \psi |^2 \Big] \;, 
\end{equation}
where $V(\psi)={\mathcal F}_0(\psi)$ is determined from the 
known value of the parallel surfaces geometry, while
$Z(\psi)$ can be computed perturbatively. Note that  $Z$ can be evaluated in practice just setting
$\eta_0=0$ in Eq.(\ref{zeta2}).

Higher orders may be derived by a natural extension of the procedure. It
also becomes apparent that for the expansion to be well-defined,  the
analytic structure of the form factors appearing in the perturbative
expansion around flat surfaces is relevant. In particular, the existence of
zero-momentum singularities can certainly render the DE non applicable; on
the other hand, this should be expected on physical grounds since those
singularities imply that the functional cannot be approximated, in
coordinate space, by the single integral of a local density. Physically, it
is a signal that the interaction becomes essentially nonlocal
(see Section~\ref{ssec:neumann}). Indeed, if written in coordinate space they would
require more than one integral over the spatial coordinates.

\subsection{Construction of the second-order Derivative Expansion}\label{ssec:imp}

We will now present a physical construction of the DE, based on a procedure
whereby one attempts to improve the PFA.

The expression for the PFA does not involve  derivatives of $\psi$, since one is using parallel 
planes to obtain the density, and
the corresponding functional is characterized by a single number, their
distance. 
The DE may then be introduced, as a rather natural way to improve the PFA,
simply by constructing an {\em improved surface density\/} ${\mathcal F}$. 
The improvement can be implemented by using the density that results from
using, at each point of the surface, a second order approximation to it.
Namely, the curved surface shall be approximated (locally) by a surface that 
makes a second order contact with it, i.e., that has the same first and
second derivatives as $\psi$. 

In other words, rather than evaluating $F$ for a constant $\psi = a$, we
shall consider evaluating it at
\begin{equation}
\psi = \psi({\mathbf y_\parallel}) \;=\; a \,+\, \eta({\mathbf y_\parallel}) 
\end{equation}
where $\eta$ is a quadratic function of its argument:
\begin{equation}
\eta({\mathbf y_\parallel}) \;=\;  \sum_{i=1}^2 b_i y_i + \frac{1}{2}
\sum_{i,j=1}^2 c_{ij} y_i y_j \;.
\end{equation}
We use ${\mathbf y_\parallel}$ for the coordinates on which $\eta$ may
depend, since we shall later one use ${\mathbf x_\parallel}$ to denote each
point on which the expansion is performed (for example, $a$ will depend on
${\mathbf x_\parallel}$). 

Since we want to construct a second-order expansion in derivatives, the
expression for $F$ will be expanded up to the second order in $b_i$ and first order in
$c_{ij}$. That expansion shall have the form:
\begin{equation}
F \;=\; F^{(0)} \,+\, F^{(1)} \,+\, F^{(2)}\,+\, \ldots 
\end{equation} 
where the index denotes the order in derivatives. 
One then has $F^{(0)}= F[a]$ (yielding the PFA contribution), while the term of
order one in derivatives is:
\begin{equation}
F^{(1)} \,=\,\int d^2{\mathbf y_\parallel} \, \Gamma_a^{(1)} ({\mathbf
y_\parallel})  \sum_{i=1}^2 b_i y_i  
\end{equation}  
with:
\begin{equation}
\Gamma_a^{(1)}({\mathbf y_\parallel}) \,=\,\Big[ \frac{\delta F[a + \eta]}{\delta \eta({\mathbf
y_\parallel})}\Big]_{\eta = 0} \;.
\end{equation}
The second order term receives two contributions:
\begin{eqnarray}\label{eq:expa1}
F^{(2)} &=& \int d^2{\mathbf y_\parallel} \, \Gamma_a^{(1)}({\mathbf
y_\parallel}) \,\big(\frac{1}{2} \sum_{i,j=1}^2 c_{ij} y_i y_j \big)
\nonumber\\
 &+& \frac{1}{2} \int d^2{\mathbf y_\parallel} \int d^2{\mathbf
y'_\parallel}\,\big(\sum_{i=1}^2 b_i y_i\big) \;
\Gamma_a^{(2)}({\mathbf y_\parallel},{\mathbf y'_\parallel}) 
\; \big(\sum_{j=1}^2 b_j y'_j \big)\;,
\end{eqnarray}
one involving the one-point kernel appearing in $F^{(1)}$, and the other
depending on the two-point function:
\begin{eqnarray}
\Gamma_a^{(2)}({\mathbf y_\parallel},{\mathbf y'_\parallel}) &=& 
\Big[\frac{\delta^2F(a + \eta)}{\delta \eta({\mathbf
y_\parallel}){\delta \eta({\mathbf y'_\parallel})}}\Big]_{\eta = 0} \;. 
\end{eqnarray}

Let is consider now the term of first order in derivatives, $F^{(1)}$.
We first see that $\Gamma_a^{(1)}$ cannot depend on ${\mathbf
y_\parallel}$ (since the functional derivative is evaluated at a constant
function), and it is therefore a constant regarding that variable.
Indeed, note that it can even be written  explicitly in terms of the density
${\mathcal F}_0(a)$, as follows:
\begin{equation}\label{eq:gammaa1}
\Gamma_a^{(1)}\;=\; \frac{1}{\mathcal S} \, \frac{\partial F[a]}{\partial
a} \;=\;\frac{\partial {\mathcal F}_0(a)}{\partial a} \;.
\end{equation}
Then we see that 
\begin{equation}
F^{(1)} \,=\, F_a^{(1)} \, \int d^2{\mathbf y_\parallel} b_i y_i \;=\; 0
\;,
\end{equation}
thus the first-order term vanishes.

Let us now consider the two contributions to the second-order term,
$F^{(2)}$, namely, $F^{(2)} = F^{(2,1)} + F^{(2,2)}$, with: 
\begin{equation}
F^{(2,1)} = \int d^2{\mathbf y_\parallel} \, \Gamma_a^{(1)}({\mathbf
y_\parallel}) \,\big(\frac{1}{2} \sum_{i,j=1}^2 c_{ij} y_i y_j \big)
\end{equation}
and
\begin{equation}
F^{(2,2)} \;=\; \frac{1}{2} \int d^2{\mathbf y_\parallel} \int d^2{\mathbf
y'_\parallel}\,\big(\sum_{i=1}^2 b_i y_i\big) \;
\Gamma_a^{(2)}({\mathbf y_\parallel},{\mathbf y'_\parallel}) 
\; \big(\sum_{j=1}^2 b_j y'_j \big)\;.
\end{equation}
Regarding $F^{(1)}$, using (\ref{eq:gammaa1}), one sees that:
\begin{equation}
F^{(2,1)} \,=\, {\mathcal S} \frac{\partial {\mathcal F}_0(a)}{\partial a}
\, \langle \eta  \rangle 
\end{equation}
where we have introduced the spatial average of $\eta$,
\begin{equation}
\langle \eta  \rangle \;\equiv\; \frac{1}{\mathcal S} \, 
\int d^2{\mathbf y_\parallel} \eta({\mathbf y_\parallel}) \;.
\end{equation}

Regarding $F^{(2,1)}$, since  $\Gamma_a^{(2)}$ can only depend on the difference between its
arguments,
\begin{equation}
F^{(2,2)} \;=\; \frac{1}{2} \sum_{i,j=1}^2 b_i b_j \, 
\int d^2{\mathbf y_\parallel} \int d^2{\mathbf y'_\parallel}
\, y_i \;\Gamma_a^{(2)}({\mathbf y_\parallel}-{\mathbf y'_\parallel}) \; y'_j
\;,
\end{equation}
and in terms of ${\widetilde \Gamma}_a^{(2)}({\mathbf k_\parallel})$, the Fourier
transform of $\Gamma_a^{(2)}$:
$$
\int d^2{\mathbf y_\parallel} \int d^2{\mathbf y'_\parallel}
\, y_i \;\Gamma_a^{(2)}({\mathbf y_\parallel}-{\mathbf y'_\parallel}) \; y'_j
\,=\, {\mathcal S} \, \lim_{{\mathbf k_\parallel}\to 0} \Big[
\frac{\partial^2{\widetilde \Gamma_a^{(2)}}({\mathbf k_\parallel})}{\partial
k_i \partial k_j}\Big] 
$$
\begin{equation}\label{eq:limit}
\,=\, {\mathcal S} \,\frac{1}{2}\; \delta_{ij} \;\lim_{{\mathbf k_\parallel}\to 0}\,\Big[
\sum_{i=1}^2 \frac{\partial^2 {\widetilde \Gamma_a^{(2)}}({\mathbf k_\parallel})}{\partial
k_i^2}\Big]
\end{equation}
where rotational invariance on the $x_3=a$ spatial planes has been used.

Thus:
\begin{equation}
F^{(2,2)} \;=\; {\mathcal S} \, Z(a) \;  \sum_{i=1}^2 b_i^2 \,
\end{equation}
with
\begin{equation}
Z(a) \,=\, \frac{1}{4}\lim_{{\mathbf k_\parallel}\to 0}\,\Big[
\sum_{j=1}^2 \frac{\partial^2}{\partial k_i^2} 
{\widetilde \Gamma_a^{(2)}}({\mathbf k_\parallel}) \Big] \;.
\end{equation}

Putting together the results for  $F^{(0)}$, $F^{(1)}$ and $F^{(2)}$ 
we see that in all of them an ${\mathcal S}$ factors out. Therefore, up to
second order in derivatives:
\begin{equation}\label{eq:second}
F[a + \eta] \,=\, {\mathcal S} \; \Big[ 
{\mathcal F}_0(a) \, + \, \frac{\partial {\mathcal F}_0(a)}{\partial a}
\, \langle \eta  \rangle \,+\, 
  \, Z(a) \,  \sum_{i=1}^2 \,b_i^2 
\Big] \;,
\end{equation}
or (since $\langle \eta \rangle$ involves two derivatives) to the same
order:
\begin{equation}\label{eq:secondb}
F[a + \eta] \,=\, {\mathcal S} \; \Big[ 
{\mathcal F}_0(a + \langle \eta \rangle )\,+\, 
 \, Z(a) \,  \sum_{i=1}^2 \,b_i^2  
\Big] \;.
\end{equation}
It then transpires that it is convenient to perform the
splitting between $a$ and $\eta$ in such a way that $\langle \eta \rangle
=0$, since then the $c_{ij}$ term  may be ignored (to the second
order). In what follows, we shall assume that that is the case.

Thus, the second order correction to the density, in the DE, is denoted by ${\mathcal F}_2$ and shall be:
\begin{equation}
{\mathcal F}_2(a,b_i,c_{ij}) \;=\;  \, Z(a) \,  \sum_{i=1}^2 \,b_i^2  
\end{equation}
and the resulting contribution to $F$ results by integrating this density,
evaluated at the proper (${\mathbf x_\parallel}$-dependent) values of $a$ and $b_i$:
\begin{equation}\label{eq:second1}
F_2 \;=\; \int d^2{\mathbf x_\parallel} \, 
 \, Z(\psi) \,   (\nabla\psi)^2 \, 
\;.
\end{equation}

The total $F$ in this second-order approximation is thus obtained by
multiplying the two contributions to the density by the area element, and integrating. This results
in $F_{DE}$, the  functional given in Eq.(\ref{eq:de}),
where, we recall, the two functions that completely determine the approximation are
obtained from the functional $F$, as follows:
\begin{equation}
V(\psi) \,\equiv \, {\mathcal F}_0(\psi({\mathbf x_\parallel})) \,=\,
\Big[{\mathcal F}_0(a)\Big]_{a = \psi({\mathbf x_\parallel})}
\end{equation}
and
\begin{equation}
Z(\psi) \,=\, \frac{1}{4}\lim_{{\mathbf k_\parallel}\to 0}\,\Big[
\sum_{j=1}^2 \frac{\partial^2}{\partial k_i^2} 
{\widetilde F_a^{(2)}}({\mathbf k_\parallel}) \Big]_{a = \psi({\mathbf
x_\parallel})} \;.
\end{equation}

Note that, from the formulae that determine the functions $V$ and $Z$, we
see that a possible practical way to obtain them is by evaluating $F(\psi)$
for $\psi = a +\eta$, with an $\eta$ which spatial average is zero, and
expanding $F$ up to the second order in $\eta$ and its derivatives.  An
important implicit assumption in the construction above is the existence of
the zero momentum limit in Eq.(\ref{eq:limit}). This is of course the same assumption
we made in the previous subsection about the analytic structure of the form factors,
that will in turn depend on
the problem being considered. We just remark once more the fact that the non
existence of that limit means that the functional $F$ shall receive
nonlocal contributions contributions beyond the PFA. In other words, it
will not be just the integral of a density. 


\subsection{Some general properties of the DE}\label{sec:properties}
We present here some consequences that follow from the form of the general
expression for the DE.

Let us first point out an important property of both the PFA and its NTLO
correction in the DE: they stem from the expansion in derivatives of a
functional, thus, they involve integrals of local functions of $\psi$ and
its derivatives. Thus, in order to assess the relative magnitude of those
terms, it is not sufficient, in general, to look at the behavior of the integrands  
at some particular points. Rather, they measure global features of the
distance function $\psi$ and its derivatives. Thus, a function may have a
small curvature radius at some region but nevertheless give rise to a small
contribution. 
Besides, note that those two terms have different properties regarding
spatial scaling. Indeed, let us consider the two terms, $F_0[\psi]$ and
$F_2[\psi]$ for a given function $\psi$, assuming that the corresponding
integrals are finite. Then introduce the scaled function 
\begin{equation}
\psi_\lambda({\mathbf x_\parallel}) \equiv \psi(\lambda {\mathbf
x_\parallel}) \:,
\end{equation}
for a scale parameter $0<\lambda<\infty$. We see, by evaluating those
terms for the scaled function, that:
\begin{equation}
F_0[\psi_\lambda] = \lambda^{-2} \, F_0[\psi] \,\,\, ;  \,\,\, F_2[\psi_\lambda] =  F_2[\psi] \;,
\end{equation} 
as it follows by performing the change of variables ${\mathbf x_\parallel}
\to \lambda^{-1} {\mathbf x_\parallel}$ in both integrals. Thus, when
the function $\psi$ has a bump or depression of characteristic size $r_0$,
the second order term will be independent of $r_0$, producing the same
contribution, even when the size goes to zero (and the corresponding
curvature diverges).

This can be illustrated by an example which appears in Casimir physics.
Assuming the boundary conditions are perfect, there is no dimensionful
object in the system except $\psi$ itself. Thus, in natural units, when
calculating the energy, both $V$ and $Z$ have to be proportional to
$1/\psi^3$. Here the role of the functional $F$ is played by the
static energy, $U[\psi]$. The zeroth and second order terms in the DE for
the energy then have the form:
\begin{eqnarray}
U_0[\psi] &=& a_0\,\int d^2{\mathbf x_\parallel} \frac{1}{\psi^3}\nonumber\\
U_2[\psi] &=& a_2\,\int d^2{\mathbf x_\parallel} \frac{|\nabla
\psi|^2}{\psi^3} \;,
\end{eqnarray}
 where $a_0$ and $a_2$ are constants.

For the revolution paraboloid $\psi = a ( 1 + \frac{{\mathbf
x_\parallel}^2}{\sigma^2})$, we may evaluate both terms exactly, what
yields:
\begin{eqnarray}
U_0[\psi] &=& \frac{\pi}{2} a_0 \, \frac{\sigma^2}{a^3} \nonumber\\
U_2[\psi] &=& 2\pi a_2 \frac{1}{a} \;.
\end{eqnarray}
Since $\sigma$ is proportional to the radius of curvature, the ratio
$\frac{\sigma^2}{a^2} >>1$, and therefore the second order term is much
smaller than the zeroth order one.

This analysis can be extended to the higher order terms. Indeed, a term
with $2 n$ ($n \in {\mathbb N}$) derivatives has in the integrand a
$1/\psi^3$ factor times a
polynomial with $2 n$ derivatives and $2 n$ powers of $\psi$ in the
numerator. Thus, under scaling:
\begin{equation}
U_{2n}[\psi_\lambda] \;=\; \lambda^{2n -2} U_{2n}[\psi] \;.
\end{equation}
Thus, for the quadratic $\psi$ that we are considering, 
\begin{equation}
U_{2n} \;\propto \; \frac{1}{a} \, (\frac{a}{\sigma})^{2n-2} \;.
\end{equation}
Of course, we have assumed the expansion to be well defined to those
orders; that is something which, as we shall see, depends on the properties
of the system.


\subsection{Fourth order term}\label{ssec:fourth}

Let us, for the sake of illustration, briefly comment here on the construction of the
fourth order term (there is no third-order one), in a similar scheme to the
one considered in~\ref{ssec:imp}.

To derive the improved $F$-density, one now has to consider:
\begin{equation}
\psi = \psi({\mathbf y_\parallel}) \;=\; a \,+\, \eta({\mathbf y_\parallel}) 
\end{equation}
where:
\begin{equation}
\eta({\mathbf y_\parallel}) \;=\;  \sum_{i=1}^2 b_i y_i + \frac{1}{2}
\sum_{i,j=1}^2 c_{ij} \, y_i y_j + \frac{1}{3!} \sum_{i,j,k=1}^2 d_{ijk}
\, y_i y_j y_k + \frac{1}{4!} \sum_{i,j,k,l=1}^2 e_{ijkl} \, y_i y_j y_k y_l \;.
\end{equation}

To find the fourth order term in derivatives, one should collect in  $F$,
terms which come from its first order contribution in $e$, the second order in $c$, the
fourth order in $b$, and also first order in $b$ and $d$, or second in $b$
and first in $c$.

Thus $F^{(4)}$ receives five different contributions, $F^{(4,j)}$ ($j = 1, ... , 5$).  
All those contributions may be evaluated in Fourier space, where
they can be expressed in terms of derivatives at zero momentum of the
corresponding functional derivative; besides, those functional derivatives,
being evaluated at $\eta=0$, are translation invariant. 

It is possible to check that  $F^{(4,1)}$ term produces a constant, and it can be ignored at this
order by a redefinition of $a$.  The remaining terms have the structure:
\begin{equation}\label{eq:ff42}
F^{(4,2)} = \frac{1}{8} \,{\mathcal S} \, \sum_{i,j,k,l}^2
\gamma^{(4,2)}_{ijkl} c_{ij} c_{kl} 
\end{equation}  
\begin{equation}\label{eq:ff43}
F^{(4,3)} = \frac{1}{24} \,{\mathcal S} \, \sum_{i,j,k,l}^2
\gamma^{(4,3)}_{ijkl} b_i b_j b_k b_l \, 
\end{equation}  
\begin{equation}\label{eq:ff44}
F^{(4,4)} = \frac{1}{6} \,{\mathcal S} \, \sum_{i,j,k,l}^2
\gamma^{(4,4)}_{ijkl} b_i b_j c_{kl} \;. 
\end{equation}  
\begin{equation}\label{eq:ff45}
F^{(4,5)} = \frac{1}{4} \,{\mathcal S} \, \sum_{i,j,k,l}^2
\gamma^{(4,5)}_{ijkl} b_i d_{jkl} \;. 
\end{equation}  
where the $\gamma$ coefficients are constant tensors, which moreover may be
further simplified by using rotational invariance. All the terms carry a
factor of ${\mathcal S}$, which appear because of momentum conservation.

We conclude by analyzing  the tensor corresponding to one of those contributions:
$\gamma^{(4,2)}_{ijkl}$. It can be obtained as follows:
\begin{equation}
\gamma^{(4,2)}_{ijkl} \,=\, \left[\frac{\partial^4
{\widetilde\Gamma}^{(2)}(p)}{\partial p_i \partial p_j
\partial p_k \partial p_l} \right]_{p \to 0} \;,
\end{equation}
and rotational invariance means that it has the form:
\begin{equation}
\gamma^{(4,2)}_{ijkl} \,=\, C \, (\delta_{ij} \delta_{kl} + \delta_{ik}
\delta_{jl} + \delta_{il} \delta_{ij} \delta_{jk} ) 
\end{equation}
where $C$ is an $a$-dependent constant, determined by the term of order four in an
expansion of the kernel at low momenta:
\begin{equation}
C = \frac{1}{8} \left[\frac{\partial^4
{\widetilde\Gamma}^{(2)}(p)}{\partial p_i \partial p_i
\partial p_j \partial p_j} \right]_{p \to 0} .
\end{equation}
When used to construct the DE to fourth order, this term shall produce:
\begin{equation}
F^{(4,2)} \,=\, \frac{1}{8} \int d^2{\mathbf x_\parallel} C(\psi) \left[ |\nabla \psi|^4
+ 2 (\partial_i \partial_j \psi)^2 \right] \;.
\end{equation}
A similar approach allows one to derive all the other contributions.

\section{Generalizations of the DA in nuclear and colloidal physics}\label{sec:other}
In this section we will discuss some generalizations of the DA proposed in
the context of nuclear and colloidal physics, from the point of view of the
DE.


\subsection{The generalized PFA in nuclear physics}

The application of the DA in nuclear physics started with a celebrated
paper by Blocki et al~\cite{Blocki1}. In that paper, the authors
rediscovered the DA in a rather different context, and applied it to compute the
interaction between nuclei. The starting point of Ref.~\cite{Blocki1} is a
Derjaguin-like formula for the interaction energy between surfaces.  That
formula incorporates, as an essential ingredient, what the authors called
`universal function' , the interaction energy between planar surfaces,
which the authors calculated using a Thomas-Fermi approximation.

To proceed, let us describe the kind of system being considered, and at the
same time introduce some notation: let us consider two surfaces $S_L$ and
$S_R$, plus an intermediate mathematical surface $S$ used to parametrize
the physical ones. For smooth and slightly curved surfaces, we expect the
interaction energy to be well-described by the PFA, as in Eq.(\ref{PFA}). 
One can now rewrite the surface integral above, by introducing the set of
level curves for $h$ in $S$: they are closed curves that correspond to a
fixed distance $h$ between $S_L$ and $S_R$.
Denoting by $J(h)dh$ the area between two curves on $S$ corresponding to
distances $h$ and $h+dh$, the PFA expression for the interaction energy $U$ can
be written as a one dimensional integral
\begin{equation}
U_{\rm PFA}=\int dh \, J(h) \, E_{\parallel}(h)\; .
\label{PFAnucJ}
\end{equation}
We then assume the surfaces to be gently curved, so that just one patch is
sufficient to describe them, and besides we use Cartesian coordinates
$(x_1,x_2)\equiv {\mathbf x_\parallel}$ on $S$. The distance between $S_L$
and $S_R$ then becomes a function $h = h({\mathbf x_\parallel})$, and $J$
is constant. 

Performing a second order Taylor expansion of $h$ around the point of
closed approach, which corresponds to a distance $d$:
\begin{equation}
h(x_1,x_2)\simeq
d+\frac{1}{2}\frac{x_1^2}{R_1}+\frac{1}{2}\frac{x_2^2}{R_2} \;,
\end{equation}
where $R_1$ and $R_2$ are the radii of curvature of the surface defined by
$x_3 = h({\mathbf x_\parallel})$, one obtains the DA of Eq.(\ref{DA}). 

In a subsequent paper~\cite{Blocki2}, a generalization of the PFA was
introduced. The starting point was again Eq.(\ref{PFAnucJ}), but now the
surfaces could have large curvatures, as long as they remained almost
parallel locally. The main difference introduced by the weaker assumptions
about the surfaces is that now the Jacobian $J$ may become a non-trivial
function of $h$, rather than being just a constant. 
 Using the linear expansion
\begin{equation}
J(h)\approx J_0+ J_1 h \;,
\end{equation}
it can be shown that the force $f$ between surfaces becomes:
\begin{equation}\label{eq:correction}
f_{\rm PFA}(d)\,=\, J_0 E_\parallel(d)- J_1(d)\int_d^\infty \, dh\,  E_{\parallel}(h) \; ,
\end{equation}
where the second term is a correction to the usual DA. Note, however, that
from a conceptual point of view, this is not a generalization of the DA,
since the starting point is the same as before: $U_{\rm PFA}$. What the
previous formula does is to provide an explicit formula for the surface
integral appearing in the PFA, which now involves a new geometrical object,
the Jacobian $J$. 

In other words, Eq.(\ref{eq:correction}) is still determined by the energy
density for parallel plates, not including corrections to that
object, like the ones appearing in the DE. That kind of
correction depends on the geometry and on the nature of the interaction.

Note that in nuclear physics there is an additional complication to deal
with: even for two infinite half-spaces separated by a gap, the interaction
energy $E_\parallel$ is not exactly known. Different approximations have
been used to compute that, and they give rise to different PFA's.  
For a recent review see Ref.~\cite{RevNuc}. 

\subsection{SEI and SIA in colloidal physics}

The methods SEI~\cite{SEI} and  SIA~\cite{SIA}, have been
introduced within the context of colloidal physics, and constitute another
generalization of the DA.  While based on different physical assumptions,
the final result in both cases is the same. 
It may be introduced as follows: consider a compact object in front of a
plane $x_3=0$, $x_3$ denoting the normal coordinate to the plane that points
towards the  compact object.

The SEI approximation to the interaction energy is then given by
\begin{equation} 
U_{\rm SEI}=-\int_{\rm plane} dx_1 dx_2\, \frac{\hat n\cdot \hat
e_3}{\vert\hat n\cdot\hat e_3\vert} \, E_\parallel\;,
\label{SEI}
\end{equation}
where $\hat n$ is the unit outward normal  to a surface element in the compact object.
When the compact object can be thought of as delimited by two surfaces, one
facing the plane and the other away from it, the SEI approximation consists
in computing the difference between the usual PFA for each surface.  

It may appear surprising, at first glance, that the two surfaces contribute
with different signs to the interaction energy. However, as we will see,
this is related to the fact that the SEI becomes exact for almost
transparent bodies, where the interaction comes from volumetric 
pairwise contributions.

In the colloidal physics literature, the SEI method is justified by assuming 
that there is a pressure on the compact object, which  should be integrated over
the closed surface  surrounding it in order to find the total force~\cite{SEI}. 
Alternatively, it has been shown that Eq.(\ref{SEI}) becomes exact when the
interaction between bodies can be obtained as the result of pair potentials
of their constituents~\cite{SIA}.  

In order to understand, and reinterpret, this formula within the context of Casimir
physics, and at the same time to provide a systematic way of evaluating the
NTLO, we shall use a rather simple example. Let us consider a quantum  scalar field
$\varphi$, in the presence of two media: one of them, denoted by $L$, 
corresponding to the $x_3 \leq 0$ half-space, while the other, $R$, is
defined in terms of two functions:
\begin{equation}\label{eq:defr}
R \;=\; \{ (x_1,x_2,x_3): \psi_1(x_1,x_2) \leq x_3 \leq  \psi_2(x_1,x_2) \}
\,,
\end{equation}
as seen in Fig. 1.

\begin{figure}
\centering
\includegraphics[width=8cm , angle=0]{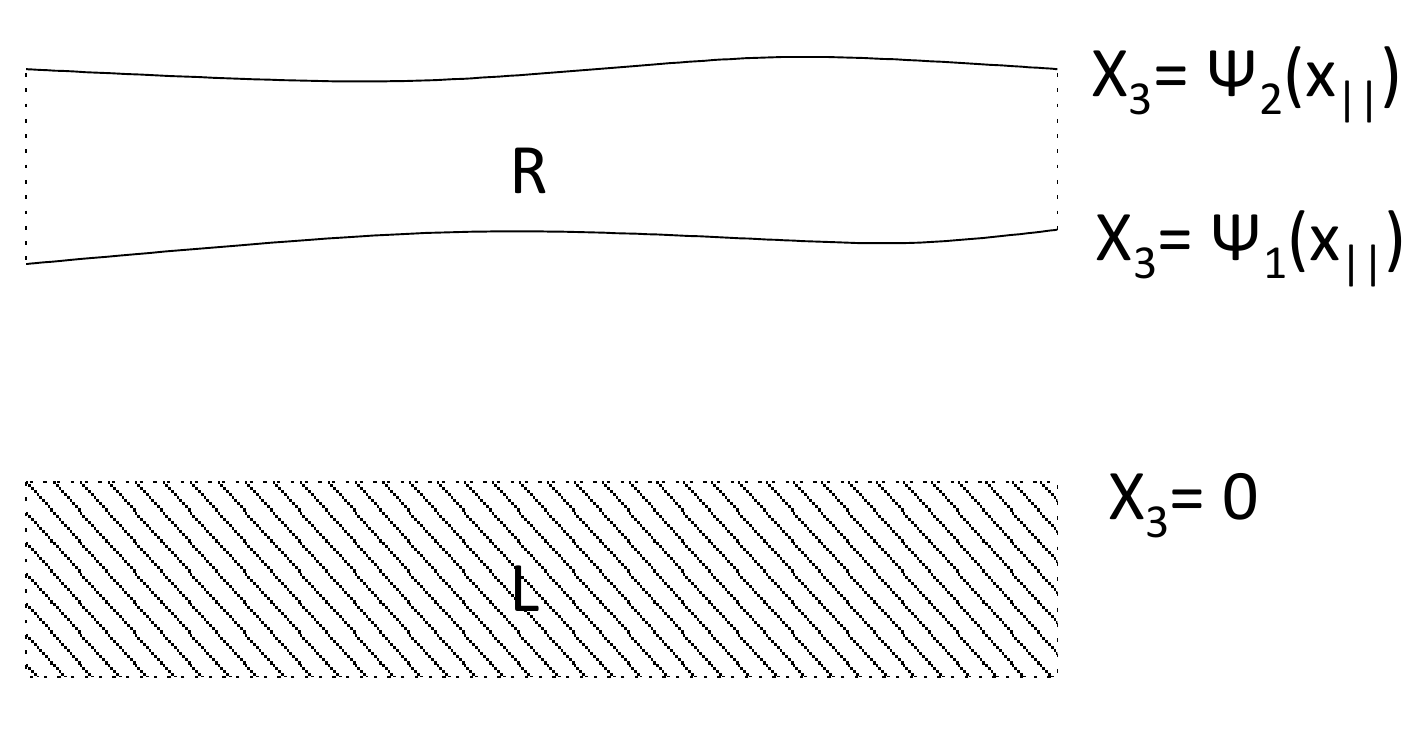}
\caption{Representation of the two media. One of them, denoted by $L$, 
corresponding to the $x_3 \leq 0$ half-space, while the other, $R$, is
defined in terms of two functions:
$R \;=\; \{ (x_1,x_2,x_3): \psi_1(x_1,x_2) \leq x_3 \leq  \psi_2(x_1,x_2) \}$} \label{fig1}
\end{figure}

Besides, we shall assume that the field propagation inside each media can be
represented by the presence of a non-vanishing interaction term.
Assuming also that outside the $L$ and $R$ regions there is vacuum, the
Euclidean action adopts the following form:
\begin{equation}
{\mathcal S}[\varphi] \;=\; {\mathcal S}_0[\varphi] + {\mathcal S}_I[\varphi] 
\end{equation}
where 
\begin{equation}
{\mathcal S}_0[\varphi] \;=\; \frac{1}{2} \int d^4x \, \big( \partial 
\varphi \big)^2 
\end{equation}
is the free action for the fluctuating vacuum field, while ${\mathcal S}_I$
contains two terms, corresponding to the $L$ and $R$ regions, respectively:
\begin{equation}
{\mathcal S}_I[\varphi] \;=\; {\mathcal S}_L[\varphi] \,+\, {\mathcal S}_R[\varphi]
\end{equation}
with:
\begin{equation}
{\mathcal S}_L[\varphi] \;=\; \frac{\lambda_L}{2} \, \int d^3x_\parallel
\int_{-\infty}^0 dx_3 \, {\mathcal L}_I(\varphi) \;,
\end{equation}
and
\begin{equation}
{\mathcal S}_R[\varphi] \;=\; \frac{\lambda_R}{2} \, \int d^3x_\parallel
\int_{\psi_1({\mathbf x_\parallel})}^{\psi_2({\mathbf x_\parallel})}  dx_3
\, {\mathcal L}_I(\varphi) \;,
\end{equation}
where $x_\parallel \equiv (x_0,x_1,x_2)$ and ${\mathcal L}_I$ is a local Lagrangian.

In what follows, we shall assume that the $R$ medium is semi-transparent,
so that the corresponding term ${\mathcal S}_R$ can be treated
perturbatively, to the first non-trivial order in $\lambda_R$. 
No assumption will be made  about the $L$ term.

The vacuum energy $U$ shall be a functional of the two functions
$\psi_{1,2}$, and may be written in terms of the vacuum amplitude
${\mathcal Z}$, as follows:
\begin{equation}
U[\psi_1, \psi_2] \;=\; - \lim_{T \to \infty} \big( \frac{1}{T} \log
{\mathcal Z} \big)\;
\end{equation}
where $T$ is the extent of the imaginary time coordinate and
\begin{equation}
{\mathcal Z} \;=\; \int {\mathcal D} \varphi \; e^{- {\mathcal S}(\varphi)}
\;.
\end{equation}
Expanding the functional integral in powers of $\lambda_R$, the lowest 
nontrivial contribution reads
\begin{equation}
U[\psi_1, \psi_2] = E[\psi_1] - E[\psi_2]\, ,
\end{equation}
with
\begin{equation}
E[\psi_{1,2}] =  \lim_{T \to \infty} 
\big[ \frac{1}{T} \langle S_{1,2}[\varphi] \rangle_L \big]\, . 
\end{equation}
Here where we have introduced
\begin{equation}
{\mathcal S}_{1,2}[\varphi] \;=\; \frac{\lambda_R}{2} \, \int d^3x_\parallel
\int_{\psi_{1,2}({\mathbf x_\parallel})}^\infty  dx_3 \, {\mathcal L}_I \;,
\end{equation}
and the $\langle \cdot \rangle_L$ symbol denotes functional average with the
weight defined by ${\mathcal S}_0 +{\mathcal S}_L $.  The crucial point is to observe that
all the dependence with the shape of the surfaces is in the lower integration limit 
of ${\mathcal S}_{1,2}$. As a consequence, we can write
\begin{equation}
U[\psi_1, \psi_2] \;=\;  \int d^2 {\mathbf
x_\parallel} \left (E_\parallel(\psi_1)-E_\parallel(\psi_2)\right)\, ,
\label{lin}
\end{equation}
where $E_\parallel(a)$ is the interaction energy per unit area between  semispaces separated by a gap
of width $a$. The physical interpretation of this result is that, being the interaction
between semispaces, in order to obtain the interaction energy for the configuration described
by $\psi_1$ and $\psi_2$, one must subtract from $E_\parallel(\psi_1)$ the contribution coming from
points with $x_3>\psi_2$. This `linearity' is valid only for the first order in $\lambda_R$.
Eq. (\ref{lin}) coincides with the result obtained using the SEI.

In order to illustrate these points, let us assume that ${\mathcal L}_I=\varphi^2$.
In the strong coupling limit for $\lambda_L$, the field satisfies Dirichlet boundary conditions at $x_3=0$.
An explicit calculation yields
\begin{equation}
U[\psi_1, \psi_2] \;=\;  \frac{\lambda_R}{32 \pi^2} \, \int d^2 {\mathbf
x_\parallel} \,\big[\frac{1}{\psi_1({\mathbf x_\parallel})} - 
\frac{1}{\psi_2({\mathbf x_\parallel})}  \big]\; ,
\end{equation}
which is the difference between the PFA energies associated to the surfaces $\psi_1$ and $\psi_2$.
It can be shown explicitly, however, that the property of the energy being
the difference between the ones corresponding to each surface is lost in
the next order in $\lambda_R$.

To summarize this section, we have shown that SEI gives an exact result for
almost transparent media, where one can use the superposition
principle and consider the interaction as the sum of pairwise potentials. 
Although this was already mentioned in Ref.\cite{SIA}, the model presented here
suggests how to go beyond the leading order, and provides a systematic way
of evaluate the interaction energy for dilute media. Finally, we would like
to mention that the fact that PFA becomes exact in this limit has also been
mentioned by authors working in Casimir physics~\cite{Decca,Miltondilute}.  

\section{Review of results}\label{sec:appl}

In this section we summarize and briefly review some of the results 
obtained through the use of the DE for the
calculation of interaction energy between surfaces, in different contexts.
It is not our intention to be exhaustive, but to enumerate some of the
results that, we believe, are noteworthy. We also take advantage to pinpoint and clarify some
aspects that deserve more comment than in their original presentation,
now under the light of this general derivation. For the Casimir effect, we consider a scalar field with Dirichlet and
Neumann conditions, at zero and finite temperature, and also an
electromagnetic field with non perfectly conducting surfaces.  We show how
the PFA emerges naturally from this approach, and also calculate the NTLO
correction to the PFA.  We also review  results obtained for the
electrostatic interaction, both for surfaces at fixed potentials and
endowed with patch potentials. 

At the end of this Section we will briefly discuss some findings related to
non analytic terms appearing in the expansion at finite temperature in the Neumann
case.

\subsection{DE for the Casimir effect}

We have shown that the PFA can be thought of as akin to the leading order
term in a derivative expansion of the Casimir energy with respect to the
shape of the surfaces in Ref.~\cite{Foscoetal2011}. Moreover, when the first non
trivial correction containing two derivatives of $\psi$ are included, the
general formula gives the NTLO correction to PFA for a general surface. The
general expression for the second-order approximation
to the interaction energy (or free energy depending the case) is the one showed in Eq.(\ref{eq:de}). 

In Ref.~\cite{Foscoetal2011} we have applied the DE to the evaluation of the
Casimir interaction energy for a scalar field with Dirichlet boundary
conditions. The calculation consisted, in terms of the general derivations
we presented in Section \ref{ssec:demo}, of the application of the expansion to an 
effective action (proportional to the energy for static boundary
conditions). The DE was obtained by performing the same calculation
suggested in Section~\ref{ssec:demo}, namely, a second-order expansion of the
functional around the parallel planes case. 

The general result for the DE approximation to the Casimir interaction energy for perfect mirrors
can be written as
\begin{equation}
U_{\rm DE}[\psi]=-\frac{\pi^2}{1440} \int d^2{\mathbf x_\parallel}\frac{1}{\psi^3}\big[ \alpha+\beta(\nabla\psi)^2\big]\, ,
\label{DE Casimir perf}
\end{equation}
where $\alpha$ and $\beta$ are numerical coefficients that depend on the field considered (scalar or electromagnetic) and 
on the boundary conditions imposed on the surfaces. This form for $U_{DE}$
can of course be anticipated by simple dimensional analysis. The zeroth order term equals the PFA approximation to 
the vacuum energy, while the second order one contains the first
non-trivial correction to PFA.

For a scalar field satisfying Dirichlet (D) boundary conditions \cite{Foscoetal2011} we have $\alpha_D=1$ and $\beta_D=2/3$.
A scalar field with Neumann (N) boundary conditions was considered  in~\cite{bimonteT},
where it was shown that $\alpha_N=1$ and $\beta_N=2/3\, (1 -
30/\pi^2)$. In the same reference, the authors presented the
results corresponding to an electromagnetic field and perfectly conducting
surfaces, which turns out to be the sum of the Dirichlet and Neumann results.

It is worth to stress the last result: within the DE approach, the electromagnetic Casimir interaction energy between
perfectly conducting surfaces  is the sum of the scalar Casimir energy for Dirichlet and Neumann boundary 
conditions. This was already known for the leading PFA approximation, and it is also valid
for the first non trivial correction \cite{nos-em}.  Of course, it will be not valid at higher orders.

Also in Ref.\cite{bimonteT},  the DE was extended to two curved surfaces,
for Dirichlet, Neumann, mixed (Dirichlet and Neumann on different surfaces)  and 
electromagnetic (perfect metal) boundary conditions. 
Ref.\cite{bimonteEPL2011} presents the leading  correction to PFA for gold
at room temperature.

Although derived for surfaces describable by a single function $\psi({\mathbf x}_\parallel)$, the DE has been applied to more general 
geometries that include compact objects in front of a plane. In these cases, the integration is restricted to a portion
of the compact object that  is closer to the plane. It has been shown that, for perfect mirrors,
the PFA and its NTLO correction are insensitive
to the choice of the integration area in the limit where the surfaces are very close \cite{Foscoetal2011}. This is not the case
for semi-transparent mirrors, as we have shown in the previous section.

In all particular examples where the NTLO correction to PFA has been computed 
analytically, the results coincide with the prediction of the derivative expansion.  This is the case for a cylinder in front 
of a plane \cite{Bordagcp} and also for a sphere in front of a plane \cite{Bordagsp}. Moreover, the DE has been useful to detect
\cite{bimonteEPL2011}
an error in previous calculations \cite{Nikolaev} of the sphere-plane interaction energy beyond PFA, that was
subsequently corrected in Ref.  \cite{Bordagsp}.  

Let us denote by $U$ the exact Casimir interaction energy for a given geometry and  by $U_{\rm PFA}$ its PFA. For both cylinder-plane and 
sphere-plane geometries the analytic NTLO correction is of the form
\begin{equation}
\frac{U}{U_{\rm PFA}}=1+\gamma \frac{a}{R}\, ,
\end{equation} 
where $a$ is the minimum distance, $R$ is the radius (of the sphere or the cylinder), and $\gamma$ a numerical coefficient that depends
on the geometry and the boundary condition. For the cylinder-plane it has been shown that
\begin{equation}
\gamma_D=\frac{7}{36}\;,\quad \gamma_N=\frac{7}{36}-\frac{40}{3\pi^2}\, ,
\end{equation}
while for the sphere-plane
\begin{equation}
\gamma_D=\frac{1}{3}\;,\quad \gamma_N=\frac{1}{3}-\frac{40}{\pi^2}\, .
\end{equation}
It has also been proved that the electromagnetic result is
the sum of the Dirichlet and Neumann cases. All these results can be reproduced using the DE by plugging the functions $\psi$ corresponding to a cylinder and 
a sphere into Eq.(\ref{DE Casimir perf}), and expanding the result of the integrals in powers of $a/R$.

The numerical calculations are also consistent with the NTLO correction
for the cylinder-plane geometry \cite{numcyl} and for the sphere-plane geometry \cite{bimonteEPL2011}, although
for Neumann boundary conditions there is a discrepancy between the analytic predictions  \cite{bimonteEPL2011, Bordagsp} 
and the numerical fit. A similar discrepancy occurs with the fit presented in Ref.\cite{Serge} for the electromagnetic case.
We believe that these discrepancies may be due to the fact that the numerical calculations have not been performed for sufficiently small values of $a/R$, 
and therefore the fits are sensitive to the particular functions and intervals used to 
obtain them. This sensitivity has been noticed in  \cite{numcyl} for the cylinder-plane geometry and in
\cite{bimonteEPL2011}, for the sphere-plane geometry. 

\subsection{Patch potentials and the electrostatic energy interaction}

In Ref.~\cite{nos_annphys}, we have applied the DE to the evaluation of the 
the electrostatic interaction energy (the functional to expand in
derivatives) between two perfectly conducting surfaces, one flat and the other slightly curved,
held  to a potential difference $V_0$.  In this situation, the interaction energy reads
\begin{equation}
\label{resannphys}
U_{\rm DE}\simeq \frac{\epsilon_0V_0^2}{2}  \int d^2{\mathbf x_\parallel} \frac{1}{\psi}\left[1+\frac{1}{3}(\nabla\psi)^2\right]\, .
\end{equation}
We have shown explicitly  that, in particular cases where analytic exact results are available, the DE reproduces
the exact  ones up to NTLO (this is the case, for instance, for a sphere or a cylinder in front of a plane).

In Ref.~\cite{patches}, we have 
extended these results to the case in which the surfaces have
patch potentials. These potentials were not introduced
as boundary conditions, but modeled  by means of electric dipole layers
that are adjacent to the surfaces. 

The result was expressed in terms of the two-point autocorrelation
functions for those patch potentials, and of the single function $\psi$ which defines the
curved surface. The reason for studying this, is based on the fact that
surface imperfections can lead to a local departure from ideal metallic
behavior, yielding space-dependent patch potential on the surface of the
mirrors. They produce a force that may be, in principle, relevant to the
interpretation of precision experiments involving two surfaces. 

In order to present a more compact expression for the results, it is
convenient to assume that the potentials' autocorrelation function depends on the variance of
the potential $V_{\rm rms}$ and on a single characteristic length ${\ell }$. 
Then, on dimensional grounds we shall have that the Fourier transform of
the auto correlation function is of the form 
\begin{equation}
{\widetilde\Omega}({\mathbf k_\parallel})=V^2_{\rm rms}\,{\ell}^2\,g(\vert{\mathbf k_\parallel}\vert{\ell}),
\end{equation}
for some dimensionless function of a dimensionless argument, $g$. 

In terms of the objects above, we have found that:
\begin{equation}
V(\psi)=\frac{V^2_{\rm rms}}{\psi}\,v({\ell}/\psi)\,\, \quad Z(\psi)=\frac{V^2_{\rm rms}}{\psi}\, z({\ell}/\psi)\, ,
\end{equation}
where
\begin{eqnarray}
v({\ell}/\psi)&=&-  \frac{2}{\pi}\frac{\ell^2}{\psi^2}\int_0^\infty dx \frac{x^2}{e^{2x}-1}g(x\ell/\psi) \nonumber\\
z({\ell}/\psi)&=&\frac{1}{16\pi}\frac{\ell^2}{\psi^2} 
\int_0^\infty
 dx\, \frac{x^2 g(x\ell/\psi) }{\sinh^5(x)}
\left [(1-8x^2)\cosh(x)-\cosh(3 x)+12 x\sinh(x)\right]\, .
\end{eqnarray}

One
can show that, when $\ell\gg \psi$, $v$ and $z$ tend to the result for constant potentials, and therefore
\begin{equation}
U_{\rm DE}\simeq-\epsilon_0 V^2_{\rm rms} \int d^2{\mathbf x_\parallel} \frac{1}{\psi}\left[1+\frac{1}{3}(\nabla\psi)^2\right]\, .
\end{equation}
This is twice the result for the electrostatic energy between surfaces held at a constant
potential difference $V_{\rm rms}$ (see Eq.(\ref{resannphys})).
The factor two comes from the fact that we are considering the
same correlation function on both surfaces~\footnote{There is also an extra minus sign, that comes from the fact that in the present calculation the potentials on the surfaces
are not produced by external batteries but are due to the internal structure of the materials.}. 

On the other hand, in the opposite limit $\ell\ll \psi$, one can make the
approximation  $g(x\psi/\ell)\simeq g(0)$ inside the integrals to get
\begin{equation}
V\simeq -\frac{g(0)\zeta(3)\ell^2}{2\pi} \frac{V^2_{\rm rms}}{\psi^3}\quad Z\simeq
-\frac{g(0)(1+6\zeta(3))\ell^2}{24\pi} \frac{V^2_{\rm rms}}{\psi^3} \;,
\label{smallarg}
\end{equation} 
which has the same dependence with distance as the Casimir energy,
something which is in this case due to the lack of a dimensionful quantity
associated to the correlation length.

\subsection{Derivative expansion at finite temperature}

In \cite{DETfinite}, we have obtained expressions for the coefficients that
determine the DE at finite temperatures, for the free energy $F$ in a Casimir
system~\footnote{We use here identical notations ($F$) for the free energy as for
the functional used in the general derivation of the DE.}.
We presented closed analytic expressions for those coefficients, in
different numbers of spatial dimensions $d$, both for the zero and high
temperature limits. We have considered surfaces satisfying both either
Dirichlet or Neumann boundary conditions, finding some qualitative differences
between those two cases:
for two Dirichlet surfaces, the NTLO term in the DE is  well
defined (local) for any temperature $T$. Besides, it interpolates smoothly
between the proper limits: namely, when  $T \to 0$ it tends to the one we
had calculated for the Casimir energy, while for $T \to \infty$ it
corresponds to the one for a $d=2$ theory, realizing the expected
dimensional reduction at high temperatures.  

The DE approach (up to second order) may be applied to this case, with the free energy as
functional of the surface. We present the Dirichlet and Neumann cases
separately. 

\subsubsection{Dirichlet boundary conditions}

In the Dirichlet case, we write the Casimir free energy as follows: 
\begin{equation}
	F_{\rm DE} [\psi] \;=\; \int d^{d-1}{\mathbf x}_\parallel \, 
\Big\{ b_0(\frac{\psi}{\beta},d) \frac{1}{[\psi({\mathbf x}_\parallel)]^d} 
\,+\, 
b_2(\frac{\psi}{\beta},d) \, \frac{(\nabla\psi)^2}{[\psi({\mathbf x}_\parallel)]^d} 
\Big\}  \label{DE dir}
\end{equation}
where the two dimensionless functions $b_0$ and $b_2$ can be obtained from
the knowledge of the Casimir free energy for small departures
around the $\psi({\mathbf x_\parallel}) = a = {\rm constant}$ case. 

In the very high (infinite) temperature limit, we have that
\begin{eqnarray}\label{eq:b0b2high}
&& \big[b_0(\xi,d)]_{\xi>>1} \;\simeq\; \xi \, \big[b_0(\xi, d-1)]_{\xi \to 0}\equiv \xi\,  b_0(d-1) \;, \nonumber\\
&& \big[b_2(\xi,d)\big]_{\xi>>1} \;\simeq\; \xi \, \big[b_2(\xi, d-1)\big]_{\xi
\to 0}\equiv \xi \, b_2(d-1)\;,
\end{eqnarray} 
where $\xi = \psi/\beta$. The coefficients $b_0(d-1)$ and $b_2(d-1)$
are those corresponding to perfect mirrors at zero temperature in
$d-1$ dimensions,  
a reflection of the well known `dimensional reduction' phenomenon at high 
temperatures, for bosonic degrees of freedom.

In particular, the DE up to the second order in the high temperature limit, in $d = 3$ dimensions, is
\begin{equation}\label{Fbetainf3}
F_{\rm DE} [\psi]\vert _{\psi/\beta >>1, d=3} \,\sim\, 
-\frac{\zeta(3)}{16\pi\beta}\int d^2{\mathbf x}_\parallel \frac{1}{[\psi({\mathbf x}_\parallel)]^{2}} \left\{ 1+ 
\frac{(1+6\zeta(3))}{12\zeta(3)} \,
(\nabla\psi)^2 \right\}  \, .
\end{equation}

Let us apply this result to the evaluation of the Dirichlet Casimir
interaction for a sphere in front of a plane. As before, we denote by $a$ the minimum distance
between the surfaces, and by $R$ the radius of the sphere.
As already mentioned, although the
surface of the sphere cannot be covered by a single function $z =
\psi({\mathbf x}_\parallel)$,  we will consider just the
region of the sphere which is closer to the plane~\cite{Foscoetal2011}.

The sphere is described by the function 
\begin{equation}
\psi = a + R \left(1 - \sqrt{1 - \frac{\rho^2}{R^2}}\right)\, ,
\label{psi sphere}
\end{equation}
where we used polar coordinates ($\rho, \phi$) for the $x_3 =0$ plane.
This function describes the hemisphere when $0\leq \rho \leq R$. The
DE will be well defined if we restrict the integrations
to the region $0\leq \rho\leq\rho_M<R$.  

We will assume that $\rho_M/R=O(1)<1$.  Inserting this expression for
$\psi$ into the free energy Eq.(\ref{Fbetainf3}), and performing explicitly the integrations
we obtain~\footnote{There is a typo in Eq.(39) of Ref.\cite{DETfinite}.}: 
\begin{equation}
F _{\rm DE} [\psi]\vert_{\psi/\beta >>1,d=3} \,\sim\, -\frac{\zeta(3)R}{8\beta a} \left(1-\frac{1}{6\zeta(3)} \frac{a}{R}\log\left(\frac{a}{R}\right)\right)\, .
\label{Fsp}
\end{equation}
Note that, as long as $a\ll R$, the force will not depend on $\rho_M$.  
As expected on dimensional grounds, the $R/a^2$ behavior of the leading contribution in 
the zero temperature case changes to $R/a\beta$ at high temperatures.  
This problem has been solved exaclty in Ref.\cite{bimonteprl}. One can readily show that
Eq. (\ref{Fsp}) coincides with the small distance expansion of the exact result.

It is interesting to remark that the  NTLO correction from the DE becomes non-analytic,
because of the integration, in the parameters defining the function
$\psi$.  This behavior has been already noted in numerical
estimations of the Casimir interaction between a sphere and a plane in the
infinite temperature limit, for the electromagnetic case in
Refs.\cite{bimonteprl,Neto2012}. Note that this non-analyticity {\em has nothing to do with the
non-analyticity of the form factors described in Section 2}.
There, the DE was not applicable, we deal here
with terms that appear in a system where the DE is perfectly well-defined. 
One integrates over the surface, and when expanding or small
$a/R$, one gets both analytic and non-analytic contributions. The
latter are not a drawback but a normal feature of the DE. 

Very recently, the free interaction energy between a sphere and a plate at high temperatures
has been computed exaclty in an arbitrary number of dimensions for Dirichlet boundary conditions \cite{teoclassical}. We have checked 
that the DE reproduces the leading and NTLO of the exact result for $d=4,5$. We sketch here
the calculations. In the high temperature limit the free energy reads
\begin{equation}
	F_{\rm DE} [\psi] \;=\;\frac{1}{\beta} \int d^{d-1}{\mathbf x}_\parallel \, 
\Big\{ b_0(d-1) \frac{1}{[\psi({\mathbf x}_\parallel)]^{d-1}} 
\,+\, 
b_2(d-1) \, \frac{(\nabla\psi)^2}{[\psi({\mathbf x}_\parallel)]^{d-1}} 
\Big\}\, .  
\label{DE dir high T}
\end{equation}
Inserting Eq.(\ref{psi sphere}) into Eq.(\ref{DE dir high T}) and expanding in powers
of $a/R$ we obtain, for $d=4$,
\begin{equation}
\frac{F_{\rm DE}}{F_{\rm PFA}}= 1+\frac{1}{4}\frac{a}{R}
\end{equation}
while for $d=5$
\begin{equation}
\frac{F_{\rm DE}}{F_{\rm PFA}}= 1+\frac{1}{3}\left(1-\frac{\zeta(3)}{\zeta(5)}\right)\frac{a}{R}\, .
\end{equation}
Both expressions are consistent  with the analytic results for the force  presented in Ref.\cite{teoclassical}.

\subsection{Neumann boundary conditions}\label{ssec:neumann}

The free energy can  be written as before (see Eq.(\ref{DE dir}), but with coefficients $c_0$ and $c_2$ instead of
$b_0$ and $b_2$.
The zero order term coincides with the one for the Dirichlet case; namely:
$c_0 = b_0$. The second-order coefficient is given by 
\begin{equation}\label{eq:defbn2}
c_2(\xi) \,=\, \frac{1}{2} \Big[\frac{\partial G^{(2)}(\xi ; n, |{\mathbf
l}_\parallel|)}{\partial |{\mathbf l}_\parallel|^2} \Big]_{n\to 0, |{\mathbf
l}_\parallel| \to 0} \;.
\end{equation} The expression of $G^{(2)}(\xi ; n, |{\mathbf
l}_\parallel|)$ has been calculated in \cite{DETfinite}. We do not present here 
the explicit expression of it since their form is not relevant for the actual presentation.  

For $d=1$, the coefficient $c_2$ coincides with its Dirichlet counterpart
$b_2$. In higher dimensions, the structure of the NTLO correction is different.

In $d=3$, the NTLO term contains, besides a standard looking local term,
also a  nonlocal contribution, linear in $T$, and therefore present for any
$T>0$. This leaves room, when the temperature is sufficiently low, to use
just the local term (of second order in derivatives) as the main correction to the PFA. 

Of course, the nonlocal term will always break down for a sufficiently high
temperature, whose value will
depend on the actual shape of the surface involved. We stress once more that this non-analytic
behavior is a consequence of the Neumann boundary conditions, and may not be present
for imperfect boundary conditions, as those considered in Ref.\cite{bimonteT}. This point deserves further
analysis.

It is important to note that a {\it local} DE breaks down for Neumann
boundary conditions at $d=2$ at zero temperature. However, one can still
perform an expansion
for smooth surfaces, including nonlocal contributions in the Casimir energy. For instance, in this 
case, the NTLO correction to the PFA will be nonlocal and  proportional to
\begin{equation}
\int d^2x_\parallel\eta(x_\parallel)\nabla_\parallel^2\log(-a^2\nabla_\parallel^2)\eta(x_\parallel)\, ,
\end{equation} where $\eta$ is a small departure from $\psi =$ const. 
The breakdown of
the local expansions is related to the existence of massless modes in the
theory. These modes are generally allowed by Neumann but not for Dirichlet
boundary conditions, that impose a mass gap of order $1/a$.

The logarithmic behavior of the form factor in $d=2$  induces a similar
non-analyticity   for $d=3$ at finite temperature.  Therefore, in an
expansion for small values of $\vert k_\parallel\vert$, in addition to  a
term proportional to $k_\parallel^2$, there is a contribution proportional
to $(Ta) k_\parallel^2
\log(k_\parallel^2a^2)$ at any non-vanishing temperature, which is not
cancelled by the rest of the sum over Matsubara frequencies.  

\section{Conclusions}\label{sec:concl}
We have presented both a construction of the DE, based on a physical
argument, and a formal derivation of it, for a general family of problems,
which can be defined in terms of a functional depending on a function
characterizing a surface as its argument.
This can be applied, as it has been done, to Casimir and electrostatic
problems. We have argued that the same procedure could also be used in nuclear and colloidal
physics, since the derivation is sufficiently general to encompass those and
other physical situations, as Casimir-like forces in critical systems \cite{crit}.  
We have made contact with the latter by comparing and putting in similar
terms the various existing PFA-like approximations, showing that they
correspond to the zeroth order in the DE.   

The existing results about the application of the DE to different contexts
have been briefly reviewed, mentioning some of the features that, we
believe, may shed new light on the respective systems.

We have shown in an explicit example, how the DE may induce non analytic
contributions in the ratio $a/R$, where $a$ is the minimum distance and $R$ 
the radius of a sphere, for the Casimir free energy between a plane and a
sphere at high temperatures. That non-analyticity is dependent on the geometry of 
the system considered, and appears in situations where the DE is well-defined. In other words, 
it is not due to the existence of non-analyticities in the momentum kernel of the second functional derivative 
of the functional $F$.

We have shown that the DE does reproduce correctly the NTLO corrections in various Casimir 
calculations. This is the case, in particular, for the sphere-plane geometry with Dirichlet boundary conditions at very high temperatures 
(the classical limit), where the result is known exactly in an arbitrary number of dimensions.

We end this paper with a few remarks about the generality of the DE approach and possible  future lines of research.
Regarding the interactions,  the DE can be applied, in principle,  both to additive and non-additive forces, 
superficial or volumetric, as long as the
interaction energy can be written as a  functional of the geometry of the surfaces.  This is the case when the surfaces 
describe homogeneous physical objects with very
small widths, or when they correspond to interfaces between different homogeneous material media.
There are of course situations where the above condition is not met. For instance,  the gravitational interaction 
between two non-homogeneous bodies cannot be described as a functional of their shapes.  

Regarding the geometry of the bodies, up to now all applications of the DE have been restricted to a particular class of geometries, i.e. surfaces 
describable by functions $x_3=\psi(x_1,x_2)$, where $x_i$ are Cartesian coordinates. It would be interesting to generalize the results to other
coordinates or, even better, to provide a covariant formulation in terms of geometric invariants of the surfaces.  Work in this direction is in progress.

\section*{Acknowledgements}
This work was supported by ANPCyT, CONICET, UBA and UNCuyo. FDM would like to thank
D. Dalvit, T. Emig, F. Intravaia, A. Lambrecht, P. Maia Neto  and S. Reynaud for discussions on this and
related matters during the 
workshop `Casimir Physics 2014'.  We would also like to thank D. Dantchev for useful comments regarding the
SEI and SIA.

\end{document}